\documentclass{ws-procs975x65}

\def\ve#1{{\mbox{\boldmath$#1$}}}
\begin{document}

\title{
TESTING LOCAL LORENTZ INVARIANCE\\ WITH HIGH-ACCURACY ASTROMETRIC OBSERVATIONS
}

\author{S. A. Klioner$^*$, S. Zschocke, M. Soffel, A. V. Butkevich}

\address{Lohrmann-Observatory, Dresden Technical University,
01062 Dresden, Germany\\
$^*$E-mail: Sergei.Klioner@tu-dresden.de}

\begin{abstract}
  This paper summarizes the analysis of the consequences of the
  violation of the Local Lorentz Invariance (LLI) on astrometric
  observations. We demonstrate that from the point of view of the LLI
  astrometric observations represent an experiment of Michelson-Morley
  type. The future high-accuracy astrometric projects (e.g., Gaia) 
  will be used to test the LLI.
\end{abstract}


\bodymatter

\section{Introduction}
\label{section-intro}

Motivated by ideas about quantum gravity, a tremendous amount of
efforts over the past decade has gone into testing Local Lorentz
Invariance (LLI) in various regimes \cite{LRR-LLI}. This paper
summarizes the framework allowing one to test LLI using
high-accuracy astrometric observations. The basic idea is that the
usual special-relativistic aberrational formulas used in the
corresponding relativistic models is a direct consequence of the Lorentz
transformations \cite{Klioner2003}. A generalization of that
aberrational formula obtained with generalized Lorentz transformation
contains parameters (similar to the Mansouri-Sexl ones) and can be
directly used to test LLI \cite{Klioner2005}. Especially the
future ESA mission Gaia \cite{Gaia} will provide a lot of
high-accuracy astrometric data that will be used to make independent
tests of LLI.

\section{Parametrized coordinate transformations}

The transformation between preferred coordinates $(T,X^a)$ and
non-preferred ones $(t, x^i)$ read \cite{MansouriSexl}:
\begin{eqnarray}
\label{t-TX}
c\,t&=&\Lambda^0_0\,c\,T+\Lambda^0_a\,X^a,
\\
\label{x-TX}
x^i&=&\Lambda^i_0\,c\,T+\Lambda^i_a\,X^a,
\end{eqnarray}
\noindent
where
\begin{eqnarray}
\label{MS_15}
\Lambda^0_{0} &=& a - b \; (\ve{\epsilon} \cdot \ve{K}) \,,
\\
\label{MS_20}
\Lambda^0_{a} &=& d \, \epsilon^a + (b - d) \frac{\ve{\epsilon} \cdot \ve{K}}
{K^2} \, K^a \,,
\\
\label{MS_25}
\Lambda^i_{0} &=& - b \, K^i\,,
\\
\label{MS_30}
\Lambda^i_{a} &=& d \delta^{i a}
+ (b - d) \frac{K^a \, K^i}{K^2}\,.
\end{eqnarray}
\noindent
Here, $\ve{K}=\ve{V}/c$, $\ve{V}$ is the velocity of 
the origin of the system $(t,x^i)$ with respect to $(T,X^a)$,
and $a$, $b$, $d$, and $\ve{\epsilon}$ are arbitrary functions of
$\ve{K}$. 

\section{Basic aberrational formulas}

Let us consider the relation between 
directions of light propagation of a given light ray 
in the preferred frame $S^a$ and that
in the non-preferred one $s^i$. Here we consider the same light ray as
seen by an observer at rest relative to $(T,X^i)$ and another observer
(co-located with the first one) at rest relative to $(t,x^i)$. Taking
the differentials along the light ray we have
\begin{equation}
S^a={1\over c}\,{dX^a\over dT},
\end{equation}
\noindent
for the preferred frame ($\ve{S}\cdot\ve{S}=1$) and
\begin{eqnarray}
p^i&=&{1\over c}\,{dx^i\over dt} \,,
\\
s^i&=&p^i/|\ve{p}|
\end{eqnarray}
\noindent
for the non-preferred frame. The last normalization is needed since the
light velocity is not equal to $c$ in the non-preferred frames and
therefore vector $p^i$ is not an Euclidean unit vector. 
Using the coordinate transformations
between $(T,X^a)$ and $(t,x^i)$ given above we get the transformations
between $\ve{S}$ and $\ve{s}$ in closed form:
\begin{eqnarray}
\label{s-S}
\ve{s}&=&
{
\displaystyle{
f\,\ve{S}+(1-f)\,{\ve{K}\,(\ve{K}\cdot\ve{S})\over K^2}-\ve{K}
}
\over
\displaystyle{
\left(
f^2+K^2-2\,\ve{K}\cdot\ve{S}+(1-f^2)\,{(\ve{K}\cdot\ve{S})^2\over K^2}
\right)^{1/2}
}
}\,,
\end{eqnarray}
\begin{eqnarray}
\label{S-s}
\ve{S}&=& \ve{K} +
\left({\left(f^2\,(\ve{K}\cdot\ve{s})^2+(1-K^2)\,\left(K^2-(\ve{K}\cdot\ve{s})^2\right)\right)}^{1/2}
- f\,K\,(\ve{K}\cdot\ve{s})\right)
\nonumber\\
&&\times{K\over K^2-(1-f^2)\,(\ve{K}\cdot\ve{s})^2}\,\left(\ve{s}-(1-f)\,{\ve{K}\,(\ve{K}\cdot\ve{s})\over K^2}\right) \,,
\end{eqnarray}
\noindent
where $f=d/b$. In the limit of special relativity one gets the normal
special-relativistic aberrational formulas. Note that the
transformation between $\ve{S}$ and $\ve{s}$ depends only on $f$ and
does not depend on $a$ and $\epsilon$. This demonstrates that the
aberrational formula tests the same properties of the Lorentz
transformation as the Michelson-Morley experiment, that is, the
isotropy of light velocity.

\section{Realistic aberrational formula}

In practice the aberrational formula entering relativistic models
\cite{Klioner2003} corrects for aberration due to the velocity of the
observer relative to the barycenter of the solar system. The solar
system barycentric reference system is not usually assumed to be
the preferred system in the sense of the LLI. Therefore, we have to
consider three reference systems: one preferred system $(T,X^a)$, two
non-preferred ones -- system $(t,x^i)$ attached to the barycenter
of the solar system, and one more system $(t^\prime,x^{i\prime})$
attached to the observer. The transformation between the preferred and
non-preferred coordinates are given above. 
The only parameter of the transformations is the velocity
of the origin of the non-preferred coordinates in the preferred ones. 
The velocity of the origin
of $(t,x^i)$ relative to $(T,X^a)$ is $\ve{V}$.
The velocity of the origin
of $(t^\prime,x^{i\prime})$ relative to $(T,X^a)$ is $\ve{V}^\prime$
and relative to $(t,x^i)$ is $\ve{v}$.  The relation between these
three velocities follows from the coordinate transformations and reads
($\ve{K}^\prime=\ve{V}^\prime/c$, $\ve{k}=\ve{v}/c$)
\begin{equation}
\ve{K}^{\prime} =
\ve{K}+ {a\over d}\,{\left(1-\ve{\epsilon}\cdot\ve{k}\right)}^{-1}\,
\left(\ve{k}-(1-f)\,{\ve{k} \cdot \ve{K} \over K^2}\,\ve{K}\right)\,.
\label{velocity_90}
\end{equation}
\noindent
Finally, denoting $\ve{s}^\prime$ 
the direction of light relative to $(t^\prime,x^{i\prime})$, 
combining (\ref{s-S})--(\ref{S-s}) written for 
two non-preferred coordinate systems and using 
(\ref{velocity_90}) one gets
\begin{eqnarray}
\ve{s}^\prime&=&\hbox{\bf P}\,\ve{s}^{\prime\prime},
\\
\ve{s}^{\prime\prime}&=&\ve{s}+(\ve{s}\cdot\ve{k})\,\ve{s}-\ve{k}
- {1\over 2}\,(\ve{s}\cdot\ve{k})\,\ve{k}
- {1\over 2}\,k^2\,\ve{s}
+ \left(\ve{s}\cdot\ve{k}\right)^2\,\ve{s}
\nonumber\\
&&
-\eta\,(\ve{s}\cdot\ve{K})\,\ve{k}
-\eta\,(\ve{s}\cdot\ve{k})\,(\ve{k}+\ve{K})
+\eta\,\left(\ve{s}\cdot\ve{k}\right)^2\,\ve{s}
+2\eta\,(\ve{s}\cdot\ve{k})\,(\ve{s}\cdot\ve{K})\,\ve{s}
\nonumber\\
&&
+{\cal O}(c^{-3})\,,
\label{final}
\end{eqnarray}
\noindent
where $\hbox{\bf P}$ is an orthogonal matrix of the Thomas-like precession,
and $\eta={1\over 2}-\beta+\delta$, where $\beta$ and $\delta$ are the
usual Mansouri-Sexl parameters ($f=d/b=1+(\eta-{1\over 2})\,K^2+{\cal
  O}(c^{-4})$, $\eta=0$ in special relativity). The Thomas precession
plays no role here since pure rotation cannot be observed in
astrometry (cannot be distinguish from the local rotation of the
observer) and the operational reference system attached to the
observer is chosen not to rotate with respect to the barycentric
reference system. Eq. (\ref{final}) gives the generalized aberrational
formula.  In addition to the barycentric velocity of the observer
$\ve{v}=c\,\ve{k}$, this formula contains the parameter $\eta$ and the
velocity of the solar system barycenter relative to the preferred
frame $\ve{V}=c\,\ve{K}$. Taking the value of $\ve{V}$ from the dipole of the
cosmic microwave background, one can determine $\eta$.  Alternatively,
both $\eta$ and $\ve{V}$ can be determined from astrometric
observations. More details on the derivation and interpretation
of this formula will be given elsewhere \cite{Klioner-future}.

\section*{Acknowledgments}

This work was partially supported by the BMWi grants 50\,QG\,0601 and
50\,QG\,0901 awarded by the Deutsche Zentrum f\"ur Luft- und Raumfahrt
e.V. (DLR).

\end{document}